\documentclass{article}
\usepackage{times}

\makeatletter

\DeclareMathAlphabet{\mathfrak}{U}{euf}{b}{n}

\def\bbbone{{\mathchoice {\rm 1\mskip-4mu l} {\rm 1\mskip-4mu l}
{\rm 1\mskip-4.5mu l} {\rm 1\mskip-5mu l}}}

\def\N{\mathsf{N}}
\def\A{\mathsf{A}}
\def\K{\mathsf{K}}
\def\SO{\mathsf{SO}}
\def\g{\mathsf{g}}
\def\ng{\mathsf{n}}
\def\kg{\mathsf{k}}
\def\ag{\mathsf{a}}

\def\so{\mathfrak{so}}
\def\k{\mathfrak{k}}
\def\n{\mathfrak{n}}
\def\a{\mathfrak{a}}

\makeatother

\begin{document}

\LARGE
\begin{center}
  {\bf  Quantum $\kappa$-Poincar\'e Algebra from de Sitter Space of Momenta}
\end{center}
\normalsize

\vspace{1ex}

\begin{center}

     Jerzy Kowalski-Glikman\footnote{E-mail: jurekk@ift.uni.wroc.pl}
and
    Sebastian Nowak\footnote{E-mail: pantera@ift.uni.wroc.pl}
\\
     Institute for Theoretical Physics, University of Wroclaw\\
     Pl. Maxa Borna 9, 50-204 Wroclaw, Pland
\end{center}

\vspace{1ex}

\begin{abstract}
There is a growing number of physical models, like point particle(s) in 2+1 gravity or Doubly Special Relativity, in which the space of momenta is curved, de Sitter space. We show that for such models the algebra of space-time symmetries possesses a natural Hopf algebra structure. It turns out that this algebra is just the quantum $\kappa$-Poincar\'e algebra.
\end{abstract}

\vspace{1ex}

\section{Introduction}

It was observed some time ago by Majid \cite{majidbook} that in phase space curvature and non-commutativity play dual role. It is well known, of course, that for curved space-time momenta do not commute. And vice versa, for more than 50 years, since the seminal work of Snyder \cite{snyder}, we know that curved momentum space implies non-commutative space-time. The importance of this duality was appreciated even more in the recent years, when the deep relation between (at least some types of) space-time non-commutativity and Hopf algebra structure of space-time symmetries has been understood. It has been shown that there exists another duality between non-commutativity of one space in the pair and non-triviality of the co-product of another space (e.g., in the case of $\kappa$-Poincar\'e algebra the non-triviality of co-product of momenta leads by Heisenberg double procedure \cite{luno} to  non-commutativity of positions.)

In particular it turned out that these duality makes it possible to construct phase space of Doubly Special Relativity (DSR). DSR has been formulated \cite{Amelino-Camelia:2000ge}, \cite{Amelino-Camelia:2000mn} as a generalization of special relativity describing the kinematics of  particles at energies close to Planck scale. This theory  possess two observer-independent scales, of velocity and mass (identified with velocity of light and Planck mass.) It soon turned out \cite{jkgminl}, \cite{rbgacjkg} that the formal structure of DSR  can be based on $\kappa$-Poincar\'e algebra \cite{kappaP}, and as the result of nontrivial Hopf structure of this algebra the space time of DSR must be non-commutative \cite{kappaM1}, \cite{kappaM2} (for reviews of recent developments in DSR see \cite{review}.)

The $\kappa$-Poincar\'e algebra is a deformed algebra of space-time symmetries, preserving the momentum scale $\kappa$. It reads\footnote{In another version of DSR theory proposed by Magueijo and Smolin \cite{Magueijo:2001cr} the algebra is different, but it can be shown to be related to the one here by change of variables \cite{juse}, \cite{Kowalski-Glikman:2002jr}. In particular the space-time non-commutativity is the same as in (\ref{I1}).}
$$
[M_i, M_j] = i\, \epsilon_{ijk} M_k, \quad [M_i, N_j] = i\,
\epsilon_{ijk} N_k,
$$
\begin{equation}\label{1}
  [N_i, N_j] = -i\, \epsilon_{ijk} M_k.
\end{equation}
$$
  [M_i, p_j] = i\, \epsilon_{ijk} p_k, \quad [M_i, p_{0}] =0
$$
\begin{equation}\label{2}
   \left[N_{i}, {p}_{j}\right] = i\,  \delta_{ij}
 \left( {\kappa\over 2 } \left(
 1 -e^{-2p_{0}/\kappa}
\right) + {{\mathbf{p}^2}\over {2\kappa}}  \right) - i\, \frac1\kappa\, p_{i}p_{j}
,\,\,\, \left[N_{i},p_{0}\right] = i\, p_{i}.
\end{equation}
It follows \cite{kappaM1}, \cite{kappaM2}, \cite{Kowalski-Glikman:2002jr} that the coordinates on dual $\kappa$-Minkowski space-time form the algebra
\begin{equation}\label{I1}
    [x_0,x_i]=-\frac{i}{\kappa}x_i
\end{equation}
as a result of a non-trivial co-product of momenta
\begin{equation}\label{3}
    \triangle (p_0)=p_0 \otimes 1 +1\otimes p_0
\end{equation}
\begin{equation}\label{4}
    \triangle (p_i)=p_i\otimes 1 +e^{-p_0/\kappa}\otimes p_i
\end{equation}

In the parallel development it was shown in \cite{Kowalski-Glikman:2002ft}, \cite{Kowalski-Glikman:2003we} that one can interpret DSR theory as a theory with curved momentum space, being de Sitter space (it turns out that Snyder space-time is also a particular instance of DSR.) Then the positions and Lorentz transformation generators algebra can be understood as the $\so(4,1)$ algebra of vectors tangent to the origin of de Sitter space.

It is interesting to note that there exist a model, in which curved space of momenta was not put by hand in one or another way, but instead derived from first principles. This model is  gravity in 2+1 dimensions coupled to point particle(s) (the clear and detailed exposition can be found in \cite{Matschull:1997du}, where the reader can also find references to the earlier papers.) It was realized recently that 2+1 gravity coupled to point particles is an example of DSR theory; the claim has been put forward that similarly in 3+1 dimensions DSR can be understood as a flat space limit of (quantum) gravity coupled to point particles; see \cite{Amelino-Camelia:2003xp}, \cite{Freidel:2003sp} for details.

In this paper we clarify the issue of relation between curved space of momenta and Hopf algebra structure of space-time symmetries. They both lead to the non-commutative structure of $\kappa$-Minkowski space-time, but the relation between them was not clear. Here we show that in the case of de Sitter space of momenta, the algebra of space-time symmetries acquires naturally the structure of Hopf algebra, and the resulting Hopf algebra is $\kappa$-Poincar\'e algebra, as expected.

\section{Iwasawa decomposition of $\so(1,4)$ algebra}

Suppose we have a theory, in which the space of momenta is four dimensional de Sitter space. As explain in Introduction we have to do with such a setting in  DSR (and in the case of 2+1 gravity coupled to a point particle, in which case de Sitter space is 3 dimensional.) The group of symmetries of this space is $\SO(1,4)$ with the algebra $\so(1,4)$. The elements of this algebra correspond to infinitesimal Lorentz transformations, which form the subalgebra $\so(1,3)$, and infinitesimal translations of momenta, which can be identified with positions.

We can uniquely decompose the algebra $\so(1,4)$ into a direct sum of subalgebras  (so called Iwasawa decomposition, see e.g., \cite{VK})
\begin{equation}\label{3.18}
    \so (1,4)=\hat{\k}+\hat{\n}+\hat{\a}
\end{equation}
 where algebra $\hat{\k}$ is the $\so(1,3)$ algebra, $\hat{\a}$ is generated by the element
 \begin{equation}\label{3.19}
    H  =\left[\begin{array}{ccccc}
    0&0 &0 &0 & 1 \\
    0 & 0 & 0 & 0 & 0 \\
    0 & 0 & 0 & 0 & 0 \\
    0 & 0 & 0 & 0 & 0 \\
    1& 0 & 0 & 0 & 0 \
  \end{array}\right]
\end{equation}
and the algebra $\hat{\n}$ by the elements
\begin{equation}\label{3.20}
  \n_i
    =\left[ \begin{array}{ccc}
 0 & \epsilon_i & 0 \\
 (\epsilon_i)^T  & 0  & -(\epsilon_i)^T \\
    0 & \epsilon_i & 0
\end{array}\right]
\end{equation}
where $\epsilon_i$ are versors in i'th direction ($\epsilon_1 = (1,0,0)$, etc), and T denotes transposition.
Every element belonging to algebra $\hat{\n}$ is a positive root of element H, whose value is equal one, i.e.,
\begin{equation}\label{3.21}
    [H,\hat{\n}]=\hat{\n}
\end{equation}
Note the similarity of this algebra and the $\kappa$-Minkowski
space-time algebra (\ref{I1}). It is easy to see that (\ref{I1})
can be obtained from (\ref{3.21}) by identification $x_0 =
-\frac{i}\kappa\, H$, $x_i = -\frac{i}\kappa\, \n_i$.

It follows that every element $\g$ of the $\SO(1,4)$ group can be decomposed as follows
 $$
 \g=(\kg\ng\ag)\quad \mbox{or} \quad \g=(\kg\,\vartheta\,\ng\ag)
$$
Here
$\kg\in  \SO(1,3)$, the element $\ag$ belongs to group  $\A$ generated by $ H$
\begin{equation}\label{3.22}
    \A=\exp\left( - \frac{p_0}{\kappa} H \right) =\left[\begin{array}{ccccc}
    \cosh \frac{p_0}{\kappa}&0 &0 &0 & -\sinh \frac{p_0}{\kappa}\\
    0 & 1 & 0 & 0 & 0 \\
    0 & 0 & 1 & 0 & 0 \\
    0 & 0 & 0 & 1 & 0 \\
    -\sinh \frac{p_0}{\kappa} & 0 & 0 & 0 & \cosh \frac{p_0}{\kappa} \
  \end{array}\right]
\end{equation}
and the element $\ng$ belongs to group $\N$ generated by algebra $\hat{\n}=p_i\, \n_i$
\begin{equation}\label{3.35}
    \N=\exp\left(\frac1\kappa\hat{n}\right)=\left[ \begin{array}{ccccc}
  1+\frac{1}{2\kappa^{2}}\vec{p}^{\;2}& \frac{p_{1}}{\kappa} & \frac{p_{2}}{\kappa}  & \frac{p_{3}}{\kappa} & - \frac{1}{2\kappa^{2}}\vec{p}^{\;2}\\
    \frac{p_{1}}{\kappa} & 0 & 0 & 0 & -\frac{p_{1}}{\kappa} \\
    \frac{p_{2}}{\kappa} & 0 & 0 & 0 & -\frac{p_{2}}{\kappa} \\
    \frac{p_{3}}{\kappa} & 0 & 0 & 0 & -\frac{p_{3}}{\kappa} \\
    \frac{1}{2\kappa^{2}}\vec{p}^{\;2}& \frac{p_{1}}{\kappa} & \frac{p_{2}}{\kappa} & \frac{p_{3}}{\kappa} & 1-\frac{1}{2\kappa^{2}}\vec{p}^{\;2} \
  \end{array}\right]
\end{equation}
and $\vartheta$=diag(-1,1,1,1,-1). Note that, as a result of isomorphism between the algebra $\hat\n+\hat\a$ (\ref{3.21}) and the algebra (\ref{I1}) the element of $\N\A$ can be equivalently expressed as an ordered plane wave on $\kappa$-Minkowski space-time
$$
\exp(ip_{i}x_{i})\exp(-ip_{0}x_{0})
$$

The decomposition of group SO(1,4) described above is unique. The coset space $\SO(1,4)/\SO(1,3)$ is de Sitter space and acting with the subgroup $\N\A$ on the stability  point ${\cal O}$
\begin{equation}\label{3.24}
 {\cal O}   =\left[\begin{array}{c}
    0 \\
    0 \\
    0 \\
    0 \\
    \kappa \
  \end{array}\right]
\end{equation}
 we get the following coordinate system
$$
 \eta_{0}=-\kappa\sinh\frac{p_{0}}{\kappa} -\frac{\vec{p}{}^{2}}{2\kappa}
 e^{\frac{p_{0}}{\kappa}}
$$
$$
 \eta_{i}=-p_{i}e^{\frac{p_{0}}{\kappa}}
$$
\begin{equation}\label{3.25}
 \eta_{4}=\kappa\cosh\frac{p_{0}}{\kappa} -\frac{\vec{p}{}^{2}}{2\kappa}
 e^{\frac{p_{0}}{\kappa}}
\end{equation}
on de Sitter space, being the four dimensional hypersurface
$$
-\eta_{0}^2 + \eta_{i}^2 + \eta_{4}^2 = \kappa^2
$$
in five dimensional space of Lorentzian signature.

It is easily seen that the coordinates $p_{0}$, $p_{i}$ describe only half of de Sitter space and therefore the group
$\K\N\A$ is not the whole group $\SO(1,4)$. However if we define
$$
 \eta_{0}=\mp (\kappa\sinh\frac{p_{0}}{\kappa} -\frac{\vec{p}{}^{2}}{2\kappa}
 e^{\frac{p_{0}}{\kappa}})
$$
$$
 \eta_{i}=\mp p_{i}e^{\frac{p_{0}}{\kappa}}
$$
\begin{equation}\label{3.29}
 \eta_{4}=\pm(\kappa\cosh\frac{p_{0}}{\kappa} -\frac{\vec{p}{}^{2}}{2\kappa}
 e^{\frac{p_{0}}{\kappa}})
\end{equation}
then the group SO(1,4) can be written in the form
\begin{equation}\label{3.30}
    \SO(1,4)=\K\N\A \cup \K\vartheta\N\A
\end{equation}
(these two parts are disjoint.) This decomposition generalizes the Iwasawa decomposition.

Thus we have the following picture: the subalgebra $\hat\n+\hat\a$ defines $\kappa$-Minkowski space-time, while energy and momenta are
just functions on the group $\N\A$ generated by the algebra $\hat\n+\hat\a$.

\section{Hopf algebra structure of momenta}

In this section we show that from the group structure we can deduce the coalgebra structure of momentum Hopf algebra.
Following the general scheme we compute coproduct, antipode and counit for momenta $p_\mu$,  which can be interpreted as functions
on the group $\N\A$.

Let $G$ be a group and let $\mathcal{A}$=Fun $G$ be  complex associative algebra of functions on G with a unit element. The
multiplication  $(f_{1},f_{2})\rightarrow f_{1}f_{2}$ and the unit $I$ in $\mathcal{A}$ are  defined
by the formula
\begin{equation}\label{3.31}
    (f_{1},f_{2})(g)=f_{1}(g)f_{2}(g), \;\;\; I(g)\equiv 1
\end{equation}
The multiplication is a mapping $\mathcal{A}\times  \mathcal{A} \rightarrow \mathcal{A}$. The algebra $\mathcal{A}$
is commutative. The group operations allow us to introduce other operations on $\mathcal{A}$, namely (see, e.g., \cite{VK})
\begin{enumerate}
\item the comultiplication $\bigtriangleup : \mathcal{A} \equiv$ Fun G $\rightarrow$ Fun (G$\times$ G),
\item the counit $\varepsilon$ : $\mathcal{A} \rightarrow$ C,
\item the antipode S: $\mathcal{A} \rightarrow \mathcal{A}$.
\end{enumerate}
 defined by the formulas
$$
(\bigtriangleup f)(g_{1},g_{2})=f(g_{1}g_{2}), \;\;\; g_{1}g_{2}\in G,
$$
$$
\varepsilon (f)=f(e),
$$
\begin{equation}\label{3.32}
    (Sf)(g)=f(g^{-1}), \;\;\; g\in G.
\end{equation}
The group properties lead to some properties of homomorphisms $\bigtriangleup$, $\varepsilon$ and $S$, to wit
\begin{equation}\label{3.33}
    (\bigtriangleup \otimes id)\circ \bigtriangleup=(id \otimes \bigtriangleup)\circ \bigtriangleup,\;\;\;\;\;
    (\varepsilon \otimes id)\circ \bigtriangleup=(id \otimes \varepsilon )\circ \bigtriangleup=id
\end{equation}
\begin{equation}\label{3.34}
    (m\circ (S\otimes id)\circ \bigtriangleup)(f)=(m\circ (id\otimes S)\circ \bigtriangleup)(f)=\varepsilon (f)I(g)
\end{equation}
It follows that algebra of functions on a group $\mathcal{A} \equiv$ Fun(G) is a Hopf algebra.
 \newline

Now we apply this formalism to the group $\N\A$ defined above. The idea is to take as the algebra of functions ${\cal A}$ on  $\N\A$, the
 momenta $p_{0}$ and $p_{i}$. As we will see, in this way we can equip the space of momenta with Hopf algebra structure. Using matrix representation of the
group $\N\A$ one can easily deduce the following property
\begin{equation}\label{3.37} \exp(-ip_{0}x_{0})\exp(ip_{i}x_{i})=\exp(ip_{i}e^{-p_{0}/\kappa}x_{i})\exp(-ip_{0}x_{0})
\end{equation}
Using this property for the product of two group elements  we get
$$    \exp(ip_{i}^{(1)}x_{i})\exp(-ip^{(1)}_{0}x_{0})\exp(ip_{i}^{(2)}x_{i})\exp(-ip^{(2)}_{0}x_{0})$$
\begin{equation}\label{3.36}
    \exp i\left(p_{i}^{(1)}+e^{-p_{0}^{(1)}/\kappa}p_{i}^{(2)}x_{i}\right)\exp -i(p_{0}^{(1)}+p_{0}^{(2)})x_0
\end{equation}
while for the inverse we have
\begin{equation}\label{3.38}    \left(\exp(ip_{i}x_{i})\exp(-ip_{0}x_{0})\right)^{-1}=\exp(-ip_{i}x_ie^{p_{0}/\kappa})\exp(ip_{0}x_{0})
\end{equation}
Comparing  these formulas with the definitions (\ref{3.32}) we  easily find that
\begin{equation}\label{3.39}
    \bigtriangleup(p_{i}) = p_{i}\otimes \bbbone +
e^{-{p_{0}/ \kappa}} \otimes p_{i}\, ,
\end{equation}
\begin{equation}\label{3.40}
    \bigtriangleup (p_{0}) = p_{0}\otimes \bbbone +  \bbbone \otimes p_{0}\, ,
\end{equation}
\begin{equation}\label{3.41}
    \varepsilon (p_{0})=\varepsilon (p_{i})=0,
\end{equation}
\begin{equation}\label{3.42}
    S(p_{0})=-p_{0}, \;\;\;\;\ S(p_{i})=-p_{i}e^{p_{0}/\kappa}
\end{equation}
We see therefore that on the algebra of momenta of DSR, being functions on de Sitter space, one can introduce the structure of Hopf algebra, and this structure is the same as in the $\kappa$-Poincar\'e algebra (\ref{3}), (\ref{4}). Let us now turn to the remaining part of $\kappa$-Poincar\'e, the deformed algebra of Lorentz symmetries.

\section{Coalgebra structure of $U(\so(1,3))$ algebra}

The coalgebra structure  of $\so(1,3)$ algebra can be deduced from the
group action. From Iwasawa decomposition described above
we know that any element of the group $\SO(1,4)$ can be uniquely decomposed in two ways
$$
\K_1\N_1\A_1=\N_2\A_2\K_2
$$
which is equivalent to
$$
\K_1\N_1\A_1\K_2^{-1}=\N_2\A_2
$$

Since this decomposition is unique the above equation defines
the action of $\SO(1,3)$ group on group $\N\A$. We can write explicitly
\begin{equation}\label{A1}
\K_1e^{i{p}_i{x}_i}e^{-ip_0x_0}\,\K_2^{-1}=e^{i{p'}_i{x}_i}e^{-ip'_0x_0}
\end{equation}
where $\K_1, \K\_2 \in SO(1,3)$ and $k_2$ depends on $p_i,p_0$ and
$\K_1$. Now if take  $\K_1-1$ infinitesimal equation (\ref{A1})
implies
\begin{equation}\label{A15}
    K_1e^{ip_ix_i}e^{-ip_0x_0}=\delta(e^{ip_ix_i}e^{-ip_0x_0})+
    e^{ip_ix_i}e^{-ip_0x_0}\,K_2
\end{equation}
If we take $K_1=1+i\xi N_i$ we get
\begin{equation}\label{A16}
    K_2=1+i\xi (e^{-{p_0}/{\kappa}}N_i
  +\frac{1}{\kappa}\epsilon_{ijk}p_jM_k)
\end{equation}
and 
$$
    \delta(e^{ip_ix_i}e^{-ip_0x_0})=
$$
\begin{equation}\label{A17}
    =e^{ip_ix_i}i\xi\left(-p_ix_0+ \left(\delta_{ij}
 \left( {\kappa\over 2 } \left(
 1 -e^{-2p_{0}/\kappa}
\right) + {{\mathbf{p}^2}\over {2\kappa}}  \right) - \,
\frac1\kappa\, p_{i}p_{j}\right)x_j\right)e^{-ip_0x_0}.
\end{equation}
Using above formulas we can write equation (\ref{A15}) in the
following form
\begin{equation}\label{A18}
  (1+i\xi N_i)e^{ip_ix_i}e^{-ip_0x_0}(1-i\xi (e^{-{p_0}/{\kappa}}N_i
  +\frac{1}{\kappa}\epsilon_{ijk}p_jM_k))\approx e^{i\mathbf{p'}\mathbf{x}}e^{-ip'_0x_0}
\end{equation}
where we remember that this equation is exact only up to linear
terms in $\xi$. Variables $p_i', p_0'$ have the following form
\begin{equation}\label{A19}
 p_j'=1-i\xi[N_i,p_j], \,\,\,\, p_0'=1-i\xi[N_i,p_0]
\end{equation}
where commutators are defined in equation (\ref{2}).

Using notation from preceding section we can write
\begin{equation}\label{A10}
    (1+i\xi N_i).f(g)=f\left((1+i\xi N_i)g\K_2^{-1}\right)
\end{equation}
where $g \in \N\A$, $f$ is a function on the group and dot means
action by commutator.

The antipode $S(N_i)$ can be defined from the action on inverse elements, to wit
\begin{equation}\label{A6}
   ( 1+i\xi N_i)\,(e^{ip_ix_i}e^{-ip_0x_0})^{-1}\K_2^{-1}=e^{i{p'}_i{x}_i}e^{-ip'_0x_0}.
\end{equation}
From the uniqueness of Iwasawa decomposition it appears  that there is
only one set $p'_0,p'_i,\K_2$ satisfying above equation. We define
the antipode $S(N_i)$ in the following way
\begin{equation}\label{A14}
    \K_2^{-1}=1-i\xi S(N_i)
\end{equation}
Using notation from preceding section and definition (\ref{A10})
we can write the left hand side of equation (\ref{A6}) in the form
\begin{equation}\label{A2}
    (1+i\xi N_i).f(g^{-1})=f\left(\left((1+i\xi S(N_i))g(1-i\xi N_i)\right)^{-1}\right)
\end{equation}
which is equivalent to
\begin{equation}\label{A3}
  (1+i\xi N_i).(Sf)(g)=(S(1+i\xi S(N_i)).f)(g)
\end{equation}
From this we can easily compute the explicit form of the antipode $S(N_i)$
\begin{equation}\label{A4}    S(N_i)=-e^{\frac{p_0}{\kappa}}(N_i-\frac{1}{\kappa}\epsilon_{ijk}p_jM_k)
\end{equation}

In order to derive the coproduct for $\so(1,3)$ algebra we must rewrite the  definition of coproduct in more convenient
form
\begin{equation}\label{A7}
    (\triangle f)(g_{1},g_{2})=\sum_i\,f_i^{(1)}\otimes f_i^{(2)}(g_1, g_2)=f(g_{1}g_{2}).
\end{equation}
where we understand above formula as multiplication
$$
\sum_i\,f_i^{(1)}\otimes f_i^{(2)}(g_1,
g_2)=\sum_i\,f_i^{(1)}(g_1) f_i^{(2)}(g_2)
$$
 We derive the coproduct of $N_i$ from
group action on product of elements belonging to group $\N\A$. This
means that we can act either on the product of translations (i.e., on $\N\A$) or
first  on translations and then take the product. We have
\begin{equation}\label{A5}   f\left(\K_1 g_1g_2\K_2^{-1}\right)=f\left(\K_1g_1h^{-1}\,hg_2\K_2^{-1}\right)=\sum_i\,f_i^{(1)}\otimes f_i^{(2)}(\K_1g_1h^{-1},hg_2\K_2^{-1}),
\end{equation}
and for $\K=1+i\xi N_i$ we get
\begin{equation}\label{A11A}
    (1+i\xi N_i).f(g_1g_2)=(1+i\xi \sum_j\,h_{j}^{(1)}\otimes h_j^{(2)})\,.\,\sum_i\,f_i^{(1)}\otimes f_i^{(2)}(g_1,g_2)
\end{equation}
where
$$
\triangle (N_i)=\sum_j\,h_{j}^{(1)}\otimes h_j^{(2)}
$$
is the Sweedler notation for coproduct. One calculates
\begin{equation}\label{A12}
  \triangle (N_i)=N_i \otimes 1 +e^{-p_0/\kappa}\otimes N_i +\frac1\kappa\, \epsilon_{ijk}p_j \otimes M_k.
\end{equation}
For the counit $\varepsilon (N_i)$ we take
\begin{equation}\label{A13}
    \varepsilon (N_i).f(g)=N_i.(\varepsilon f)(g)\; \Rightarrow \; \varepsilon (N_i)=0
\end{equation}
This whole procedure can be repeated for generators of rotation $M_i$, and the result is
\begin{equation}
   \triangle (M_i)= M_i \otimes 1 + 1 \otimes M_i, \quad S(M_i) = -M_i, \quad \varepsilon (M_i)=0.
\end{equation}

From the group properties one can deduce the corresponding properties for antipode counit and coproduct (see (\ref{3.33})), which are necessary to define Hopf algebra.

\section{Conclusions}

In this paper we show that when the phase space has the space of momenta of the form of de Sitter space, in this space one can quite naturally introduce the structure of Hopf algebra. This algebra can be further extended to  $\kappa$-Poincar\'e algebra of all ten space-time symmetries.

This result can be rather easily extended to the case when the momentum part of the phase space is anti-de Sitter space (as in the case of 2+1 gravity, as described in \cite{Matschull:1997du} and DSR with space-like deformation \cite{Blaut:2003wg}), and, most likely, to the case of arbitrary symmetric space. The fact that in these cases the algebra of space-time symmetries has  the structure of Hopf algebra may have profound physical consequences, which should be closely investigated.

It should be also noted that it follows from the investigations presented here, that similar, though, in a sense, upside-down structure arises in theories on de Sitter space-time (though in this case there would be a nontrivial Hopf algebra of positions and Lorentz transformations.) It is not unlikely that this observation may shed some new light on quantum field theory on de Sitter space, which is a basis of inflationary cosmology.

\end{document}